\documentclass[a4paper,11pt]{article}
\pdfoutput=1 

\usepackage{jheppub}
                     
\usepackage[T1]{fontenc}  
\graphicspath{{Figures/}} 
\usepackage{booktabs}     
\usepackage{bbold}        
\usepackage{mathtools}
\usepackage{multirow}
\usepackage{physics}
\usepackage[symbol]{footmisc}
\usepackage{cancel}
\usepackage{dsfont}

\graphicspath{{Figures/}} 	     

\DeclareMathOperator{\GeV}{\text{GeV}}
\DeclareMathOperator{\TeV}{\text{TeV}}

\DeclareMathOperator{\Z2}{\mathbb{Z}_2}
\DeclareMathOperator{\SMgroup}{SU(3)_c\times SU(2)_L \times U(1)_Y}

\newcommand{\Treh}{T_{\text{reh}}}

\title{\bf Baryogenesis and Dark Matter in the Mirror Twin Higgs}

\author[a,b]{Pedro Bittar,} 
\author[a]{Gustavo Burdman}
\author[a]{and Larissa Kiriliuk}
\affiliation[a]{Department of Mathematical Physics,
Institute of Physics, 
University of São Paulo, \\
R. do Matão 1371, São Paulo, 
SP 05508-090, Brazil \vspace{0.2cm}}

\affiliation[b]{Department of Physics and Enrico Fermi Institute, University of Chicago, \\
933 East 56th Street, Chicago, Illinois 60637, U.S.A.}

\emailAdd{bittar@if.usp.br}
\emailAdd{gaburdman@usp.br}
\emailAdd{larissa\_kiriliuk@usp.br}

\abstract{We consider a natural asymmetric dark matter (ADM) model in the mirror twin Higgs (MTH). We show that it is possible to obtain the correct dark matter (DM) abundance when a twin baryon is the DM without the need of explicit breaking of the MTH $\Z2$ symmetry in the dimensionless couplings (i.e. without hard $\Z2$ breaking). We illustrate how this is possible in a specific baryogenesis setup, which also leads to ADM. In the simplest scenario we obtain $m_{\rm DM}\sim O(1)$~GeV, just above the proton mass. We show estimates for direct detection rates at present and future experiments.}

\begin{document} 

\maketitle
\flushbottom

\section{Introduction}
\label{sec:intro}
The standard model (SM) of particle physics is an extremely successful quantum field theory describing the interactions of all known elementary particles.\footnote{The only exception, gravity, is non-renormalizable, and its effects can be safely neglected up to extremely high energies.} Nonetheless, there remain many questions that need to be addressed by the SM. Among them is the nature of dark matter, the origin of the baryon asymmetry, and the stability and origin of the only energy scale appearing in the SM. The Mirror Twin Higgs Model (MTH)~\cite{Chacko:2005pe,Chacko:2005un,Chacko:2005vw}, originally conceived to stabilize the electroweak scale, can be an intriguing source of dark matter candidates. For instance,  Refs.~\cite{GarciaGarcia:2015fol}  and \cite{Craig:2015xla} consider thermal relics in the MTH and  fraternal~\cite{Craig:2015pha} twin scenarios, respectively. The possibility of Asymmetric  Dark Matter (ADM) \cite{Petraki:2013wwa,Zurek:2013wia} in TH models is considered in  Refs.~\cite{GarciaGarcia:2015pnn,Terning:2019hgj} for the fraternal TH, and in  Refs.~\cite{Farina:2015uea,Farina:2016ndq} for a variety of Twin Higgs scenarios,  but most importantly for us, in the context of the MTH.

In the MTH~\cite{Chacko:2005pe}, the SM is extended to have a twin SM copy supplemented by a $\Z2$ symmetry. The Higgs sector realizes the spontaneous breaking of a global symmetry at some scale $f$. The SM Higgs is then a pseudo-Nambu-Goldsone boson of this breaking, explaining the stability of the weak scale $v$, at least up to the energy scale $\simeq 4\pi f$. Experimental bounds, mostly from the unobserved invisible Higgs boson decays to the twin sector, impose the need for a {\em soft} $\Z2$ breaking\footnote{In some cases it is even possible to have the MTH with an exact $\Z2$ symmetry as shown in \cite{Csaki:2019qgb,Beauchesne:2015lva,Yu:2016bku,Batell:2019ptb,Jung:2019fsp,Yu:2016swa}}, resulting in $f/v > 1$. However, whatever the origin of this soft $\Z2$ breaking, this does not reintroduce the hierarchy problem since the $\Z2$ breaking is assumed to be valid in the ultra-violet (UV).

In Ref.~\cite{Farina:2015uea}, the MTH  was considered to build a model for DM. There it is argued that if a twin baryon is to provide the correct DM abundance, it is necessary to introduce a {\em hard} $\Z2$ breaking in order to allow for
\begin{equation}
  m_{\rm DM} \simeq 5\,m_N~,
  \label{mdm5}
\end{equation}   
where $m_N$ is the nucleon mass and $m_{\rm DM}$ is the mass of the twin baryon.  The need for hard $\Z2$ breaking results from the fact that just using the soft breaking (i.e., $f/v>1$)  is not enough to obtain the desired value in (\ref{mdm5}). Its effects in the renormalization group running in the twin sector, appearing through the modification of quark masses and the resulting speed up of the twin QCD running, results in only a mild enhancement of $\tilde\Lambda_{\rm   QCD}$, much smaller than the factor of $5$ needed. Thus, the introduction of hard breaking in the twin QCD coupling. Although it is possible to arrange for the hard breaking to be small enough not to reintroduce the hierarchy problem, this remains an ad hoc aspect of the ADM models in the MTH.

In this paper, we consider a MTH scenario with an additional sector responsible for the baryon and dark matter asymmetries. As we show below, one of the features of the model is that it allows for the correct DM abundance even if (\ref{mdm5}) is not satisfied therefore vacating the need for hard $\Z2$ breaking. This is achieved by showing that we can obtain different baryon and DM {\em number densities}  without such breaking. In this way, in order to  obtain the correct DM to baryon abundance ratio
 \begin{equation}
\frac{\Omega_{\rm DM}}{\Omega_B} = \frac{n_{\rm
    DM}}{n_B}\,\frac{m_{DM}}{m_N}, 
  \label{eq:ratioab1}
\end{equation}
with $m_{\rm DM}\sim O(1) m_N$, the ratio of number densities must be different. The models we consider require the addition of a sector resulting in baryon number violation on both sides of the MTH. We illustrate this with simple models of baryon number violation with out-of-equilibrium decays. These models are mostly available in the literature as applied to the SM alone. We aim to show the general mechanism that allows ADM models in the context of the MTH to obtain the correct DM abundance without hard $\Z2$ breaking. On the other hand, it is interesting that the resulting models address the hierarchy problem, the origin of dark matter, and the baryon asymmetry in a natural way.

The rest of the paper is organized as follows: In the next section, we review the status of ADM models in the context of the MTH. In Section~\ref{sec:baryonadm}, we propose mechanisms for generating both the baryon and dark matter number densities needed on both sides of the MTH without incurring hard $\Z2$ breaking. Finally, we conclude in Section~\ref{sec:conc}.

\section{Asymmetric Dark Matter and the Mirror Twin Higgs} 
\label{sec:admth}

The Twin Higgs mechanism \cite{Chacko:2005pe,Chacko:2005un,Chacko:2005vw} was originally introduced as a possible solution to the little hierarchy problem. A copy of the SM matter and interactions, supplemented by a $\Z2$ symmetry, results in a global symmetry ($SU(4)$) which is spontaneously broken at a scale $f$, resulting in a spectrum of Nambu-Goldstone bosons that make up the SM-like Higgs doublet. The SM interactions explicitly break the global symmetry generating a Higgs potential and leading to electroweak symmetry breaking. In the original version, which we call the MTH, all SM particles and interactions are mirrored in the twin sector. However, as it was first pointed out in Ref.~\cite{Craig:2015pha}, the minimum matter content in the twin sector that addresses the little hierarchy problem does not require an entire copy of the SM but just a twin third generation. This case is the so-called fraternal TH (FTH).  

The twin Higgs scenario provides several possibilities for DM model building. For instance, models with thermal relics have been considered in the context of the FTH in Refs. \cite{GarciaGarcia:2015fol,Craig:2015xla}. In these cases, the twin tau is cosmologically stable due to an accidental $U(1)$ lepton number. The \textit{WIMP miracle} is recreated since these DM candidates have masses of tens of $GeV$ up to about $100$ ~GeV, and the twin weak interactions determine their thermal relic abundance. Also, in the FTH case, Ref.~\cite{GarciaGarcia:2015pnn} examines asymmetric DM (ADM) models. The preferred scenario involves a light twin b quark with a mass below $\tilde \Lambda_{QCD}$, asymmetry connected to the SM baryon asymmetry through some UV mechanism, and a cosmologically long-lived twin b baryon. One of the main advantages of the FTH scenarios is that they minimize the new relativistic degrees of freedom, which makes it easier for them to avoid conflicts with the cosmological bounds on $N_{\rm eff.}$. On the other hand, as we will see below, it is more natural to build ADM models in the MTH scenario.

 We consider the MTH model with an effective cutoff of $\Lambda\simeq  4\pi f$, where $f$ is the spontaneous symmetry breaking scale of the twin Higgs global symmetry. In the limit of exact $\Z2$ symmetry,  $f=v$, with $v$ the vacuum expectation value of the Higgs doublet in the  SM sector. However, the current experimental bounds from the measurements of the Higgs boson couplings at the LHC \cite{ATLAS:2021vrm,CMS:2022dwd} require the $f/v \gtrsim 3$~\cite{Burdman:2014zta}. This requirement can be achieved by assuming a {\em soft breaking} of the $\Z2$ symmetry, i.e., a breaking occurring in the infrared (IR) by some mechanism that respects the $\Z2$ symmetry in the ultraviolet (UV). This soft breaking guarantees that the hierarchy problem is not reintroduced in loops correcting the Higgs potential since the UV $\Z2$ symmetry forces the cancellation of contributions quadratically dependent on the cutoff in the Higgs boson two-point function. The soft $\Z2$ breaking paradigm can accommodate all known collider phenomenology with minimal tuning~\cite{Burdman:2014zta}.

The twin sector of the MTH is particularly well suited to building models of dark matter. In particular, here we consider the scenario where twin baryons, which carry an accidentally conserved global charge just as protons carry baryon number, may constitute all of the observed DM abundance. Thus, we focus on ADM models in the context of the MTH, in which the origin of the baryon and twin baryon asymmetries are related and at the heart of the apparent similarity in the DM and baryon abundances. In particular, it was shown in Ref.~\cite{Farina:2015uea} that the twin neutron in the MTH model is a viable candidate for DM. This results from a scenario where twin neutrinos somehow acquire large masses in order to avoid tight constraints from the cosmological measurements of $N_{\rm eff.}$. However, the twin photon is still in the spectrum. If only twin baryon number $\tilde B$ is generated in the twin sector (i.e., no twin lepton number $\tilde L$), then charge neutrality of the universe results in the generation of a net twin neutron $\tilde n$ number after the twin QCD phase transition. Although $\tilde{\pi}^\pm$ are also stable, their abundance is negligible~\cite{Farina:2015uea}, whereas $\tilde{\pi}^0$ still decays to twin photons. Finally, nucleosynthesis does not proceed without light twin neutrinos, and we conclude that DM is made entirely of $\tilde n$.

On the other hand, Ref.~\cite{Farina:2015uea} also raised a problem with this picture. If the softly broken $\Z2$ implies that the number densities of baryon and DM are similar, i.e.
\begin{equation}
  n_B\simeq n_{DM}~,
  \label{eqnumberds}
  \end{equation} 
then \eqref{eq:ratioab1} implies \eqref{mdm5}. However, it seems that in order to achieve $m_{\rm DM} \simeq 5 m_B$, the $\Z2$ symmetry has to be broken in the UV. To see this, we notice that in this scenario
\begin{equation}
  m_{\rm DM} \sim \tilde{\Lambda}_{\rm QCD}~,
  \label{twinqcdscale}
\end{equation}
where $\tilde{\Lambda}_{\rm QCD}$ is the twin sector strong interaction IR scale. But if we only allow for a soft $\Z2$ breaking, the only effects raising this scale compared to $\Lambda_{\rm   QCD}$ are given by the enhancements of the twin quark masses. This results in a speed-up of the running giving 
\begin{equation}
  \tilde{\Lambda}_{\rm QCD} \simeq 1.4 \,\Lambda_{\rm QCD}~,
\label{softz2break}
\end{equation}
slightly depending on the value of $f/v$. In this way, with only a soft $\Z2$, we have
\begin{equation}
  m_{\rm DM} \simeq O(1)\,m_N~.
  \label{eq:mdmo1}
\end{equation}
Thus, if the baryon and twin baryon number densities, $n_B$ and $n_{\rm DM}$ in \eqref{eq:ratioab1},  were to be equal, we could not obtain the correct DM abundance.

Ref.~\cite{Farina:2015uea} argues that the $\Z2$ symmetry forces the number density equality and that, in order to obtain the correct DM abundance, the only way out is a {\em hard breaking} of the $\Z2$ symmetry, which would be enough to give $m_{\rm DM} \simeq 5\, m_N$. These masses can be achieved, for instance, by having different values of the QCD and twin QCD couplings at the cutoff $\Lambda$. However, this reintroduces two-loop contributions to the Higgs mass squared that are quadratic in $\Lambda$. It was then argued that it is possible to introduce enough $\tilde\alpha(\Lambda) - \alpha(\Lambda)$ to obtain the desired value of $\tilde\Lambda_{\rm QCD}\simeq 5\Lambda_{\rm QCD}$ at the same time that a fine-tuning of at the most $\simeq 1\%$ is required.  

The situation described above, although technically feasible, is far from satisfactory. The MTH remains a natural extension of the SM controlling the Higgs mass UV sensitivity even after the most recent LHC bounds~\cite{ATLAS:2021vrm,CMS:2022dwd}, which come mostly from the constraints on the Higgs couplings, but also from the invisible Higgs boson branching ratio~\cite{ATLAS:2022yvh,CMS:2022qva}. Therefore it is desirable to maintain this feature of the model, i.e., to avoid forcing the MTH scenario into a fine-tuned corner of parameter space for the purpose of obtaining the correct DM abundance. Luckily, as we will show below, it is possible to avoid introducing a hard breaking of the $\Z2$ and still obtain the observed DM abundance. The key point, of course, is to relax the approximate equality $n_{\rm DM} \simeq n_B$ to accommodate eqns. (\ref{mdm5}) and \eqref{eq:ratioab1} while still using the result \eqref{softz2break}, i.e. without introducing {\em ad hoc} hard $\Z2$ breaking. 

Although we present a full model in Section~\ref{sec:baryonadm} as proof of principle for how this can be achieved, we can sketch the general idea here, independently of the specific baryogenesis model chosen to be used both in the SM and twin sectors. 
The specific baryogenesis model will set  $n_B$  after annihilation of the symmetric part of the particle-antiparticle plasma, leaving an asymmetric component. The final number density depends on the CP asymmetry,  $\epsilon_{CP}$~\cite{Kolb:1990vq}. The details of its computation depend on the specific  model under consideration. However, $\epsilon_{CP}$ is proportional to the relative complex phases of the couplings of the theory, denoted here as $\sin\phi$. On the other hand, we assume that in the twin sector DM is generated by the twin version of the same mechanism, i.e. is asymmetric. Thus, the same structure for the final number density appears in the DM sector of the MTH. The remaining asymmetric component of the DM plasma sets the final DM number density \cite{Petraki:2013wwa,Kaplan:2009ag,Shelton:2010ta,Elor:2022hpa}. Therefore, the twin $\widetilde{\epsilon}_{CP}$ will also be proportional to the complex phase of the couplings of the twin sector, which we generically denote as $\sin\widetilde\phi$. Due to the $\Z2$ symmetry, the baryon and DM asymmetries have the same microscopic origin; however, if the baryon and DM phases $\phi$ and $\widetilde\phi$ are different, we can rewrite \eqref{eq:ratioab1} as
\begin{equation}
    \frac{\Omega_{\rm DM}}{\Omega_{B}} \sim \frac{m_{\rm DM}}{m_N}\bigg|\frac{\sin\widetilde\phi}{\sin\phi}\bigg|\simeq 5.
    \label{eq:Abundance_ratio_5}
\end{equation}
From \eqref{eq:Abundance_ratio_5}, we can see that it is possible to satisfy the DM abundance ratio to baryons if there is an order one misalignment between the phases $\phi$ in the visible sector and $\widetilde\phi$ in the twin sector. 
We argue that this can be the case even in the absence of $\Z2$ breaking in the UV, given that the relative phases in the visible and twin sectors maybe be defined by IR processes that are not necessarily  identical and, therefore, could generally come from a {\em soft} $\Z2$ breaking. A simple example is the vacuum alignment leading to the spontaneous breaking of the twin global symmetry at the scale $f$. This can be compared with the vacuum alignment in the visible sector leading to electroweak symmetry breaking at the scale $v$. There is no reason why the visible sector vacuum expectation value (VEV) should be real relative to the twin sector VEV.  
 We can parameterize the vev of the Higgs bi-doublet as 
\begin{equation}
    \langle H\rangle =f 
    \begin{pmatrix}
        0\\
        \sin\theta\\
        0\\
        e^{i\delta}\cos\theta
    \end{pmatrix}.
    \label{eq:twin_VEV}
\end{equation}
where the Twin Higgs VEV $\left \langle H_B \right \rangle$ has a relative phase $\delta$ with respect of the SM Higgs VEV $\left \langle H_A \right \rangle$. This relative phase propagates to the couplings of visible and twin sector states, e.g. fermions, coupled to the SM and twin Higgses. As a result, the SM Higgs couplings will have relative phases with respect to their twin sector counterparts. For instance, this implies that the CKM phases in the twin sector in the MTH need not be the same as those in the SM. Furthermore, since we are introducing couplings among quarks and new fields in both baryogenesis and darkogenesis, the
overall relative phase of these is potentially receiving additional IR misalignment.  
We note that the phase $\delta$ in (\ref{eq:twin_VEV}) does not have physical consequences in the coupling of the Higgs boson to SM gauge bosons and fermions. This can be seen by writing the Higgs doublets in the unitary gauge~\cite{Burdman:2014zta}
\begin{equation}
    H_A = \left(\begin{array}{c}0\\f\sin{\left(\frac{v+h}{\sqrt{2}f}\right)}
    \end{array}\right) \qquad \quad
  H_B = \left(\begin{array}{c}0\\e^{i\delta}f\cos{\left(\frac{v+h}{\sqrt{2}f}\right)}
    \end{array}\right)   ~,
\end{equation}
where $h$ is the physical Higgs boson. The phase $\delta$ however, will not impact the couplings of the Higgs to gauge bosons, which can be extracted from 
\begin{equation}
    |D_\mu^A H_A|^2 + |D_\mu^B H_B|^2~,
\end{equation}
and will not depend on $\delta$.
Similarly, the couplings to fermions will result in
\begin{equation}
    \lambda_t H q_A t_A + \lambda_t e^{i\delta}\left(f-\frac{1}{2f} H^\dagger H\right)q_B t_B~,
    \label{eq:Higgs_top_cs}
\end{equation}
where we illustrate the point with the top quark couplings, and  $H$ is the SM Higgs doublet. First, we see from the expression above that the phase $\delta$ only affects the Higgs couplings to the twin (invisible) sector. 
Second, we can see that the cancellation of the quadratic divergences induced by the top quark is still efficiently performed by the twin top sector and it is not spoiled by the phase $\delta$. This is because the twin top loop contribution to the Higgs boson two point function from the second term in (\ref{eq:Higgs_top_cs}) requires the mass insertion $\lambda_t\,e^{i\delta} f$ as well as the non-renormalizable twin top-Higgs coupling which, in order to close the loop, enters like $\lambda^*_t\,e^{-i\delta}$.
This clearly shows that this relative phase between the two sectors does not spoil the UV cancellation and, therefore does not affect the $\Z2$ symmetry in the UV, so it can be considered a purely IR or soft $\Z2$ breaking.

Thus, we conclude that the phases entering in (\ref{eq:Abundance_ratio_5}) need not be related by a $\Z2$ transformation in the UV and can differ by order one values, which may result in the correct DM abundance even if the ratio of DM to nucleon masses is still just over unity. In the next section we show an explicit model of ADM in the MTH in which this mechanism is successfully implemented.

\section{Baryogenesis and Darkogenesis}
\label{sec:baryonadm}
In this section, we specify a model of baryogenesis and its MTH counterpart to exemplify that it is possible to obtain a successful ADM dark matter abundance in this context without introducing hard $\Z2$ breaking. 

\subsection{Baryogenesis}
\label{sec:Bgen}

We start by providing a simple and concrete model for baryogenesis. We generate the baryon asymmetry directly at low scales below the sphaleron decoupling temperature. Baryogenesis at low temperatures is appropriate for the twin Higgs since we expect the theory to be completed at the UV scale $\Lambda=4\pi f \approx 10 \TeV$. Therefore, we can imagine that the UV completion can play a role in the baryon asymmetry generation.

The model is based on the out-of-equilibrium decays of a singlet fermion $N_\alpha$ that violates baryon number. The need for CP-violation requires at least two flavors of $N_{1,2}$, with a mass hierarchy $M_{N_2}>M_{N_1}$, so that the tree-level and loop amplitudes can interfere with different phases. We also require the existence of a colored scalar $X$ in the $\mathbf{(3,1)_{2/3}}$ representation of the $\SMgroup$ group. We then add the following interactions to the SM sector:
\begin{equation}
\Delta\mathcal{L}_{\text{Bgen}}= \lambda_{i\alpha} N_\alpha \bar{X}^a (u_R^{~i})_a + \xi_{ij} \epsilon^{abc} X_a ({d_R^{~i}})_b ({d_R^{~j}})_c + h.c.
\label{eq:NdecaysBgen}
\end{equation}   
Here, $i,j$ are the quark generation indices, $\alpha=1,2$ is the neutral fermion flavor, and $a,b,c$ are color indices. Because of the antisymmetric nature of $\epsilon^{abc}$, the $\xi_{ij}$ coupling must be antisymmetric in flavor. This model is often considered in the context of low-temperature baryogenesis \cite{Davidson:2000dw,Babu:2006xc} since it is a simple realization of baryon number violation without proton decay \citep{Allahverdi:2010im,Davoudiasl:2010am,Cui:2012jh,Arnold:2012sd,Allahverdi:2013mza,Cheung:2013hza,Reece:2015lch,Assad:2017iib,Fornal:2020poq}.

The baryon asymmetry is generated by the decay of the lightest neutral fermion, $N_1$. The baryon asymmetry parameter is given by 
\begin{equation}
Y_{\Delta B}=\frac{n_{N_1}}{s}\left(\frac{\Gamma(N_1\rightarrow B) - \Gamma(N_1\rightarrow \bar B)}{\Gamma (N_1\rightarrow \text{tot})}\right) \equiv Y_{N_1}  \epsilon_{CP}^{N_1},
\label{eq:Nasym}
\end{equation}
where $\Gamma(N\rightarrow f)$ are the decay widths of $N_1$ to baryon number $B=+1$ or $B=-1$ final states, $Y_{N_1}\equiv \tfrac{n_{N_1}}{s}$ is the $N_1$ yield, $s$ is the entropy density,  and $\epsilon_{CP}^{N_1}$ is the CP asymmetry. We proceed to compute each piece of \eqref{eq:Nasym} separately.

\begin{figure}
\centering
\includegraphics[scale=0.06]{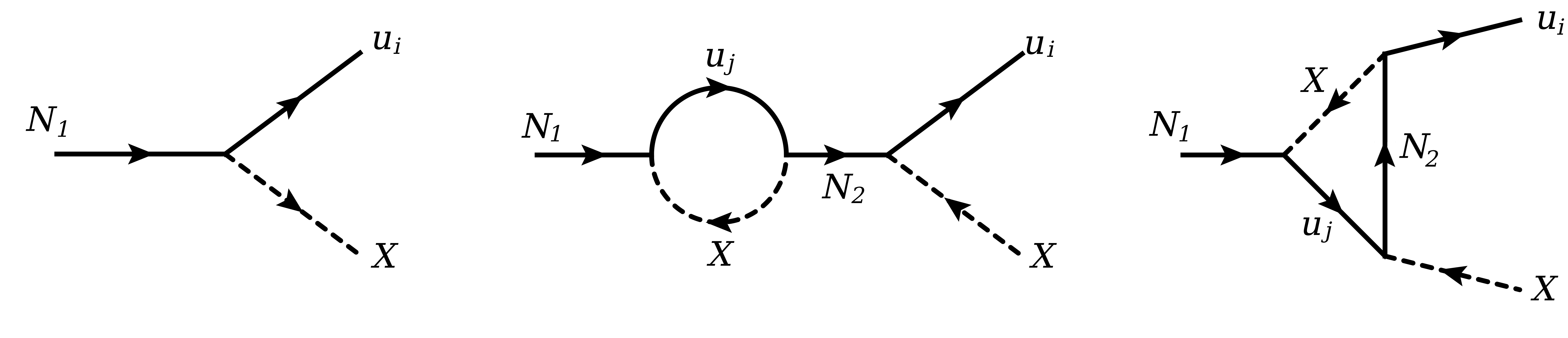}
\caption{CP violating decays of $N_1$ responsible for generating the baryon asymmetry.}
\label{fig:CPdecay}
\end{figure}

We start with $\epsilon_{CP}^{N_1}$. CP violation results from the interference of tree-level and loop amplitudes in $N_1$ decay as indicated in Fig.~\ref{fig:CPdecay}. The decay amplitude can be written as 
\begin{equation}
	\mathcal{M} = c_0 \mathcal{A}_0 + c_1 \mathcal{A}_1,
\end{equation}
where in $c_0$ and $c_1$, we separate all the couplings in the matrix elements. We can then write the CP asymmetry in the generic form~\cite{Davidson:2008bu}
\begin{equation}
	\label{eq:CPpar}
	{\epsilon}_{CP}^{N_1}= \frac{\Gamma (N_{1}\rightarrow Xu_{i}^{c})-\Gamma (N_{1}\rightarrow \overline{Xu_{i}^{c}})}{\Gamma (N_{1}\rightarrow Xu_{i}^{c})+\Gamma (N_{1}\rightarrow \overline{Xu_{i}^{c}})}=\frac{{\rm Im}\{c_0c_1^{*}\}}{\sum_\alpha|c_0|^2}
 \frac{2\int{\rm Im}\{  {\cal A}_0 {\cal A}_1^{*} \}  \tilde{\delta}~d\Pi_{uX} }
{ \int |{\cal A}_0|^2 \tilde{\delta}~d\Pi_{uX} }\, ,
\end{equation}
where $\tilde{\delta}=(2\pi)^4\delta^4(p_i-p_f)$ for the initial and final state momenta and $d\Pi_{uX}$ is the final state phase space factor.

From \eqref{eq:CPpar} we see that to have a CP asymmetry, there must be a complex phase in the product of the couplings as well as a non-zero relative phase of the two matrix elements,  ${\cal{A}}_0 {\cal{A}}_1$. To meet this last condition, we need on-shell intermediate states in the loop diagrams so that their matrix elements have a complex phase relative to the tree level one. This, in turn,  imposes a lower bound on the mass of $N_1$ from the mass of the colored scalar $X$, which should be above the $\TeV$ scale due to LHC bounds,
\begin{equation}
	M_{N_1}>M_X \gtrsim \text{few $\TeV$}.
\end{equation}
Assuming that the couplings are complex \footnote{There is no reason a priori for the couplings to be real. In the following sections, we comment on possible sources for the complex phases.}. Using the diagrams in Fig.~\ref{fig:CPdecay}, we obtain
\begin{equation}
	\label{eq:epsCP}
	\epsilon_{CP}^{N_1} =\frac{\sum_{i,j }\textup{Im}(\lambda _{i1 }\lambda _{i2 }^{*}\lambda _{j2 }^{*}\lambda _{j1 })}{24\pi \sum_{i}|\lambda _{i1 }|^2}\left [ 3 \mathcal{F}_{S}\left ( \frac{M_{N_2}^{2}}{M_{N_1}^{2}} \right )+\mathcal{F}_{V}\left ( \frac{M_{N_2}^{2}}{M_{N_1}^{2}} \right )\right ],
\end{equation}
where the functions  $\mathcal{F}_{S,V}(x)$ coming from the loop diagrams are defined as 
\begin{equation}
	\label{eq:loopF}
	\mathcal{F}_{S}(x) =\frac{2\sqrt{x}}{x-1},\hspace{10mm}  \mathcal{F}_{V}(x)=\sqrt{x}\; \textup{ln}\left ( 1+\frac{1}{x} \right ).
\end{equation}
Assuming that there are no flavor hierarchies between the $\lambda_{i\alpha}$ couplings, we take  $\lambda_{i\alpha}=\lambda_{u\alpha}$ for all $i=1,2,3$ as a simplifying approximation. Then the CP asymmetry is
given by
\begin{align}
\epsilon_{CP}^{N_1} 	&= \frac{3|\lambda _{u2}|^2 }{24\pi} \left [ 3 \mathcal{F}_{S}\left ( \frac{M_{N_2}^{2}}{M_{N_1}^{2}} \right )+\mathcal{F}_{V}\left ( \frac{M_{N_2}^{2}}{M_{N_1}^{2}} \right )\right ] \sin\phi,
    \label{eq:CPexpanded}
\end{align}
where $\phi$ is the complex phase of the product of the couplings $\lambda _{u1}\lambda _{u2}^{*}\lambda _{u2}^{*}\lambda _{u1}$. 

\begin{figure}
\centering
\includegraphics[scale=0.06]{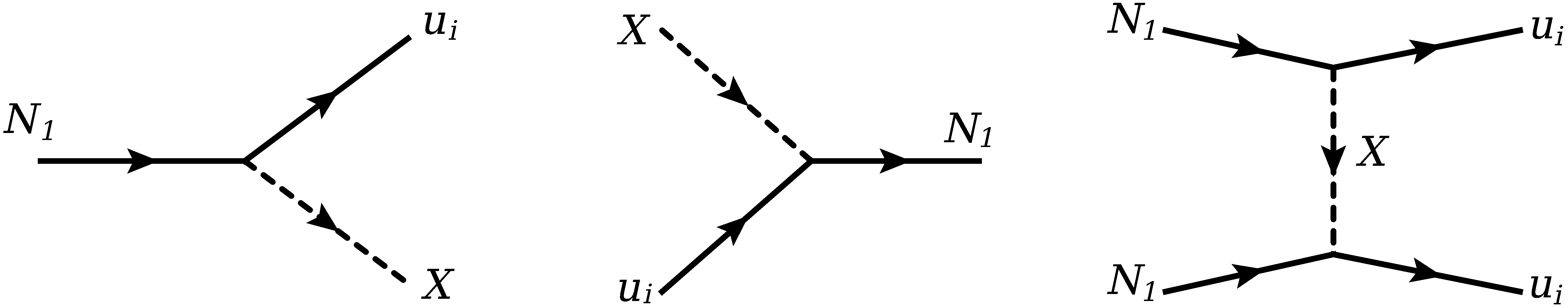}
\caption{Decay, inverse decay, and annihilation processes in the early plasma for the baryogenesis model we consider.}
\label{fig:N1_processes}
\end{figure}

Next, we turn our attention to the $N_1$ yield $Y_{N_1}$. Before the decay of $N1$, the yield will be constant once the early universe processes cease due to the expansion rate. $Y_{N_1}$ is set either thermally or non-thermally depending on the physical processes at play and the values of couplings and masses. The important scale that distinguishes the two cases is $T_{FO}$, the freeze-out temperature of the processes that change the number density of $N_1$ in the thermal bath. In this case, the relevant processes are the decay of $N_1$, its inverse decay, and $N_1$ annihilations as shown in Fig.~\ref{fig:N1_processes}.

Once the freeze-out conditions are satisfied, the inverse decay and annihilation processes will stop happening, leading to the following conditions,
\begin{equation}
\left.\frac{\Gamma_{u X \rightarrow N_1}}{2H}\right|_{T^{\rm inv.}_{FO}}=1,  \mspace{60mu} \left.\frac{\Gamma_{N_1 N_1\rightarrow u u}}{2H}\right|_{T^{\rm ann.}_{FO}}=1.
\end{equation}
In these, $\Gamma_{i}$ are the process rates for the two reactions , and $T^{\rm inv.}_{FO}$ and $T^{\rm ann.}_{FO}$ are their respective freeze out temperatures. Since we will be working with small couplings, we can safely assume that the freeze-out temperatures of these reactions will be high, above the mass of $N_1$. Therefore, $N_1$ is produced in equilibrium at high energies in the thermal scenario, and the thermal distribution determines its yield. As the inverse decay and annihilations freeze out, only the decay process will change the number density of $N_1$, mainly after the lifetime of $N_1$ has elapsed.

As argued before, the out-of-equilibrium decays of $N_1$ are responsible for the CP and baryon number violation required for baryogenesis. In addition, we need to compute the $N_1$ lifetime to know when baryogenesis mostly occurs. A long enough lifetime is required to reach the post-sphaleron baryogenesis window, as inverse decays and annihilations usually freeze out at higher temperatures. The lifetime of $N_1$ is given by 
\begin{equation}
    \tau_{N_1} = \frac{1}{\Gamma_{N_1\rightarrow u X}}= \frac{16\pi^2}{3 |\lambda_{u1}|^2}\frac{m_{N_1}^3}{m_{N_1}^4-m_X^4},
\end{equation}
where we included all the decay channels for different flavors of $u_{i=\{u,c,t\}}$. Assuming that $N_1$ and $X$ have masses above the few-$\TeV$ region\footnote{In this model, $N_1$ and $X$ can be significantly heavier than the few-$\TeV$s without changing the mechanism and the coupling bounds we derived.}, we place a bound on the coupling $\lambda_{u1}$ by requiring the lifetime to be larger than the cosmological time for sphaleron processes on the one hand and lower than the time of BBN. This results in
\begin{equation}
    \big(\tau_{\text{BBN}} \approx 10 ~s) \geq \tau_{N_1} \geq  \big(\tau_{\text{Sph}}\approx 10^{-12} ~s\big) \hspace{0.2cm}\Rightarrow \hspace{0.2cm} 10^{-14} \lesssim \lambda_{u1} \lesssim 10^{-7}
    \label{eq:N1_lifetime}
\end{equation}

To compute the thermal yield, we assume that the early universe processes decouple relativistically as it leads to the largest value of $Y_{N_1}$. As we will see shortly, this requirement will minimize the necessary tuning between the different couplings $\lambda_{u1}$ and $\lambda_{u2}$ when fixing the observed value for the baryon asymmetry. For relativistic freeze-out, the equilibrium distribution gives the following yield
\begin{equation}
    Y_{N_1}=Y_{N_1}^{EQ}\simeq\frac{45\,\zeta(3)}{2\pi^4}\frac{g_N}{g_{*,S}(T)},
    \label{eq:Yield}
\end{equation}
where $g_N$ is the $N_1$ effective number of internal degrees of freedom, and $g_{*,S}$ is the total entropic effective number of degrees of freedom\footnote{The number of effective degrees of freedom is approximately twice the SM since we need to include the twin states as they are coupled to the SM bath through the Higgs portal.}.

Finally, we impose the correct baryon asymmetry in order to constrain the parameters of the theory. Using \eqref{eq:Nasym}, \eqref{eq:Yield} and \eqref{eq:CPexpanded} we have
\begin{equation}
    Y_{\Delta B}=\frac{45\,\zeta(3)}{16\pi^5}\frac{g_N}{g_{*,S}(T)}|\lambda _{u2}|^2 \left [ 3 \mathcal{F}_{S}\left ( \frac{M_{N_2}^{2}}{M_{N_1}^{2}} \right )+\mathcal{F}_{V}\left ( \frac{M_{N_2}^{2}}{M_{N_1}^{2}} \right )\right ] \sin\phi.
    \label{eq:Yield_2}
\end{equation}
The observed baryon abundance measured by the \textit{Planck} telescope \cite{Planck:2018vyg} is $Y_{\Delta B}^{exp}=(8.75\pm 0.23)\times 10^{-11}$. Then, we can write \eqref{eq:Yield_2} as
\begin{equation}
    \frac{Y_{\Delta B}}{8.7\times 10^{-11}}=\left(\frac{213.5}{g_{*,S}(T_{FO})}\right) \left(\frac{\scriptstyle 3 \mathcal{F}_{S}\left( 1.5 \right) +\mathcal{F}_{V}\left( 1.5 \right)}{15.3}\right) \left(\frac{|\lambda _{u2}|\sin^{1/2}\phi}{2.3\times 10^{-4}}\right)^2 .
    \label{eq:Yield_parametric}
\end{equation}
We have selected $m_{N_2}^2=1.5~ m_{N_1}^2$ as a benchmark point, yet the final abundance only exhibits a weak dependence on the specific choice of mass splitting between the neutral fermions. For larger splittings, the value of the coupling $|\lambda_{u2}|\sin^{1/2}\phi$ is expected become slightly larger.

As we will argue later, we do not have any theoretical information on the origin of $\phi$, the relative phase in the couplings. Because of this, we can assume the phases have values uniformly distributed from $0$ to $2\pi$\footnote{Note that this range means that the phase $\sin\phi$ can be negative. However, we can redefine what particle or anti-particle means in this case, thus always making the baryon asymmetry parameter positive.}. If this is the case, the quantity $\sin\phi$ is naturally expected to be an order $\mathcal{O}(1)$ parameter. Therefore, we see from \eqref{eq:Yield_parametric} that we can fix the observed value for the baryon asymmetry assuming that the coupling of $N_2$ is of order $\lambda_{u2}\sim 10^{-4}$. This value means that in this specific model of low-temperature baryogenesis, a coupling hierarchy between $\lambda_{u1}$ and $\lambda_{u2}$ is necessary. The coupling $\lambda_{u1}$ must be smaller in order for $N_1$ to be long-lived enough to get to post-sphaleron temperatures, and $\lambda_{u2}$ must be larger to reproduce the observed value of the baryon asymmetry, i.e.
\begin{align}
     \label{eq:lambda_1_hierarchy}      \quad 10^{-14} \lesssim &|\lambda_{u1}| \lesssim 10^{-7},\\
    \label{eq:lambda_2_hierarchy}      & |\lambda_{u2}|\sim 10^{-4}.
\end{align}
Although the couplings above are required to be rather small, these values are radiatively stable. For instance, we can estimate the corrections to $\lambda_{u1}$ by looking at the UV contributions to the loop diagrams in Figure~\ref{fig:CPdecay}. The first diagram gives a contribution approximately given by
\begin{equation}
\delta\lambda_{u1}    \simeq \frac{\lambda_{u1}}{16\pi^2}\,\lambda_{u2}^2\,\frac{\Lambda}{m_{N_2}} 
\end{equation}
where $\Lambda$ denotes the cutoff. 
Since we assume that the states $N_1$, $N_2$ and $X$ are associated with the UV sector completing the twin Higgs model and this is not specified, the cutoff to be used in estimating these corrections should be $\Lambda\simeq 4\pi f$. Then we would have 
\begin{equation}
\delta\lambda_{u1}    \simeq \frac{\lambda_{u1}}{4\pi}\,\lambda_{u2}^2\,\frac{f}{m_{N_2}}~, 
\end{equation}
which is still quite small even for the smallest values of $m_{N_2}$.
The last diagram in Figure~\ref{fig:CPdecay} gives a milder logarithmic cutoff dependence with similar coupling factors. 
A similar argument can be made for the coupling $\lambda_{u2}$ with the appropriate replacements. Thus, although to have a fundamental understanding of the sizes of $\lambda_{u1}$ and $\lambda_{u2}$ we would need to know the details of the UV completion above $\Lambda$, these values of the couplings are radiatively stable in the context of this small extension of the twin Higgs model.

On the other hand, this hierarchy of couplings is only necessary if we assume thermal production of $N_1$. Conversely, $N_{1}$ could be non-thermally produced if an additional mechanism was active after the freeze-out temperature of the processes in Fig.~\ref{fig:N1_processes}. One possibility is the production via the decay of a new heavy particle, for example, the "reaheaton" $\tau$, a scalar field that induces a reheating period in early cosmology. The reheaton could be originated in different BSM scenarios, like non-thermal DM sectors, inflationary models, or SUSY/string models for the early universe \cite{Allahverdi:2010xz,Amin:2014eta,Allahverdi:2010im,Reece:2015lch,Douglas:2006es,Kawasaki:1995cy,Watson:2009hw,Allahverdi:2021grt}. The specific origin of this particle is beyond the scope of our work. The advantage of the $N_1$ non-thermal production mechanism is that there is no need for a hierarchy between the couplings $\lambda_{u1}$ and $\lambda_{u2}$, such as the one in \eqref{eq:lambda_1_hierarchy} and \eqref{eq:lambda_2_hierarchy} for the thermal case. We may then consider, for simplicity, that both couplings are equal in absolute value and that $N_{1}$ decays shortly after the reaheton decay. The prompt decay of $N_1$ is achieved by a larger $\lambda_{u1}$ coupling, of order $\lambda_{u1}\sim 10^{-4}$. 
In this case, the overall baryon asymmetry is given by 
\begin{align}
    Y_{\Delta B} =  \frac{n_\tau}{s}\left(\frac{\Gamma(\tau\rightarrow N_1 \rightarrow B)-\Gamma(\tau\rightarrow N_1 \rightarrow \overline{B})}{\Gamma(\tau\rightarrow \text{tot})}\right) =Y_{\tau}\: \text{Br}_{N_{1}}\: \epsilon_{N_{1}}^{CP}
    \label{eq:cptau}
\end{align}
Where $\epsilon_{N_{1}}^{CP}$ is given by \eqref{eq:epsCP}, $Y_{\tau}$ is the non-thermal yield due to the decay of the reheaton, and $\text{Br}_{N_1}$ is the branching ratio of the reheaton decay into $N_1$.

We can estimate the non-thermal yield as a function of the reheating temperature by calculating the number density at this earlier matter-radiation equality epoch. The non-relativistic energy density of the reheaton-dominated universe is $\rho\simeq m_\tau n_\tau$. At the radiation epoch, we have the energy density, $\rho_\tau=\tfrac{\pi^2}{30}g_*(T) T^4$. Therefore, we have $n_\tau \sim \tfrac{\Treh^4}{m_\tau} $ and we can obtain the reheaton yield as
\begin{align}
    Y_{\tau} = \frac{n_{\tau}}{s_{RH}} \simeq \frac{3}{4}\frac{\Treh}{m_\tau}.
    \label{eq:reh_yield}
\end{align}
Here, $s_{RH}=\tfrac{2\pi^2}{45}g_{*,S}(\Treh)\Treh^3$ is the entropy at the reheating temperature. We also used that the effective degrees of freedom $g_*(T)$ and $g_{*,S}(T)$ are approximately equal at this early epochs. The reheating temperature can be estimated by the freeze out of the reheaton processes in the early universe. The reheaton decay rate can be obtained assuming that since it is a long-lived particle, its decay is possibly mediated by a nonrenormalizable operator. For example, in  Refs.~\cite{Allahverdi:2010im,Allahverdi:2010xz,Reece:2015lch} a dimension five operator was used, resulting in
\begin{equation}
    \Gamma_{\tau}=\frac{\alpha^2}{2\pi}\frac{m_\tau^3}{M_*^2}
\end{equation}
where $\alpha$ is the effective coupling of the processes and $M_*$ is some high scale of the theory. Then, comparing with the Hubble rate at the freeze-out temperature, we have
\begin{equation}
    \Treh\simeq \frac{\Gamma_\tau^{1/2} M_{Pl}^{1/2}}{g_{*}^{4}(T)}=\frac{\alpha m_\tau^{3/2}}{(2\pi)^{1/2}g_*^4(T)}\frac{M_{Pl}}{M_*^2}.
\end{equation}
Then, assuming $M_*\sim M_{Pl}$, the reheaton yield \eqref{eq:reh_yield} is
\begin{equation}
    Y_\tau\simeq \frac{3}{4} \frac{\alpha}{(2\pi)^{1/2}g_*^4(T)}\left(\frac{m_\tau}{M_{Pl}}\right)^{1/2}
\end{equation}
With equation \eqref{eq:Yield_parametric} and \eqref{eq:cptau}, we see that $Y_{\tau} \lesssim 10^{-3}$. 
Notice that the reheating temperatures are below the range of temperatures in which spharaleons processes are active. Because of this, the couplings $\lambda_{u1}$ can be larger. Consequently, $N_{1}$ does not need a long lifetime, and baryogenesis will occur at lower temperatures, close to the reheating temperature. 

Parametrically, we can write the baryon asymmetry for the non-thermal case as
\begin{equation}
\frac{Y_{\Delta B}}{8.7\times 10^{-11}}=\left (  \frac{213.5}{g_{*}(T_{\textup{reh}})}  \right )^{4}\left ( \frac{m_{\tau}}{100\textup{TeV}} \right )^{1/2}\left(\frac{\scriptstyle 3 \mathcal{F}_{S}\left( 1.5 \right) +\mathcal{F}_{V}\left( 1.5 \right)}{15.3}\right) \left(\frac{\alpha^{1/2} |\lambda _{u2}|\sin^{1/2}\phi}{1.1\times 10^{4}}\right)^2 .
    \label{eq:Y_nonthermal}
\end{equation}

Notice that in the non-thermal case, the effective degrees of freedom $g_{*}(T_{\textup{reh}})$ can assume values with different orders of magnitude depending on the chosen reheating temperature. The reheaton may decay at any time between post-sphaleron and prior to the BBN time. In the numerical example above, we chose a reheating temperature before the SM and Twin QCD phase transition, which yields a total effective degrees of freedom of $g_{*}\simeq 213.5$. 

To conclude this section, we emphasize that our proposal is independent of the specific details of the visible sector baryogenesis model. Reproducing the observed baryon asymmetry is sufficient for the purpose of this work. Our primary focus will be to demonstrate how the $\Z2$ symmetry ensures the origin of the dark matter abundance in the twin sector, with a particular focus on the baryon asymmetry dependence on the phase and coupling given either by \eqref{eq:Yield_2} or \eqref{eq:Y_nonthermal}. What is clear is that a generic baryogenesis model that relies on the out-of-equilibrium decay of some new particle should have a similar dependence on these parameters. As such, our findings have implications beyond the specific model we have presented. We leave the extension of the model to higher temperatures via leptogenesis, the achievement of baryogenesis without hierarchical couplings, and the origin of the reheaton in the non-thermal case for future work that focuses explicitly on baryogenesis.

\subsection{Twin Darkogenesis}

Now that we have a successful baryogenesis model, we can compute the corresponding DM asymmetry using the twin mechanism discussed in Section 2. The $\Z2$ mirror symmetry results in a twin baryon asymmetry that generates the dark matter abundance in the twin sector. Then, to compute the DM abundance, we use the baryogenesis model of Section \ref{sec:Bgen}. This approach means that, analogously to the SM sector case, we introduce an out-of-equilibrium, CP, and baryon number violating decay. The new twin sector particles are two neutral fermions $\tilde{N}_{1,2}$ and a twin-colored scalar $\tilde X$ in the $\mathbf{(3,1)_{2/3}}$ representation of twin QCD, $\tilde{SU}(3)$. Then, we can add the following interactions to the twin part of the theory,
\begin{equation}
\Delta\mathcal{L}_{\text{Bgen}}^{\text{twin}}= \widetilde{\lambda}_{i\alpha} \widetilde N_\alpha \overline{\widetilde X}_a ({\widetilde{u}_{R}})^i_a + \widetilde\xi_{ij} \epsilon^{abc} \widetilde{X}_a ({\widetilde{d}_R})^i_b ({\widetilde{d}_R})^j_c + h.c.
\label{eq:Twin_lag}
\end{equation}   
Here, $i,j$ are the twin-quark generation indices, $\alpha=1,2$ is the neutral fermion flavor, and $a,b,c$ are twin-color indices. The tilde superscripts indicate that all quantities are associated with the twin sector. 

The mechanism for generating the ADM abundance is the result of imposing the $\Z2$ symmetry on the baryogenesis mechanism of the previous section. The out-of-equilibrium decay of the twin $\tilde{N}_1$ violates CP and baryon number and generates a CP asymmetry. As for the case of the $N_1$, here, the $\tilde{N}_1$ yield can be obtained thermally or non-thermally. Then, we can write the DM asymmetry as the twin baryon asymmetry parameter as
\begin{align}
    &\frac{Y_{DM}^{\rm thermal}}{8.7\times 10^{-11}}\,\,=\left(\frac{213.5}{g_{*,S}(T_{FO})}\right) \left(\frac{\scriptstyle 3 \mathcal{F}_{S}\left( 1.5 \right) +\mathcal{F}_{V}\left( 1.5 \right)}{15.3}\right) \left(\frac{|\widetilde{\lambda} _{u2}|\sin^{1/2}\widetilde{\phi}}{2.3\times 10^{-4}}\right)^2,
\end{align}
where the expression above corresponds to the thermal determination of the yield. In the non-thermal process, we would have a similar expression for the DM asymmetry, except for the non-thermal yield \eqref{eq:reh_yield}. 

Our work does not introduce any new interactions between the visible and twin sectors generating the baryon and DN asymmetries. In this way, we emphasize that the mirror $\Z2$ mechanism uniquely gives their common origin without any need for cogenesis and asymmetry transfer between the two sectors. Thus, this method for realizing the ADM idea is similar to previous models of mirror DM \cite{Petraki:2013wwa,Kobzarev:1966qya,Foot:1991bp,Hodges:1993yb,Berezhiani:2000gw,Foot:2003jt,Foot:2014mia}. One could worry that additional renormalizable interactions could be present or generated in the theory. However, these are very suppressed in our model. Since the particles we introduced in the SM sector do not interact with the Higgs directly and the Higgs portal is the only communication between the twin and SM sectors, any visible-twin interactions happen only at multiple loop order. Lastly, the MTH is expected to be UV completed near the cutoff of the theory. Therefore, one could imagine that both $\Delta\mathcal{L}_{\text{Bgen}}$ and $\Delta\mathcal{L}_{\text{Bgen}}^{\text{twin}}$ have a common origin approximately at the scale $4\pi f$. In this case, there could be more renormalizable portals beyond the twin Higgs. However, to introduce any new portals, we would need to make assumptions about the structure of the UV theory, which is beyond the scope of this paper. The most important aspect of the baryogenesis extensions we added to both sectors is that they leave the hierarchy problem unaffected, as there are no new interactions with the Higgs.

Since the DM abundance is larger than the baryon abundance, some source of misalignment will be necessary to achieve darkogenesis. The relation between the abundances is 
\begin{align}
    \frac{\Omega_{DM}}{\Omega_{B}}= \frac{n_{DM}}{n_{B}}\frac{m_{DM}}{m_{B}}\sim 5.
    \label{eq:omegaDm_omegaB}
\end{align}
As discussed in Section \ref{sec:admth}, if there is a process that enforces $n_{B}\sim n_{DM}$, we must have $m_{DM} \sim  5m_{B}$. It is difficult to achieve this mass around $5\GeV$ with only soft $\Z2$ breaking in the MTH. To see this, we observe that given that DM is a twin nucleon, the ratio of the QCD and twin QCD confinement scales can predict the ratio between the DM and nucleon masses. In appendix \ref{app:QCD_twinQCD}, we derive the ratio of the two QCD scales, which is given by 
\begin{align}
    \frac{\Tilde{\Lambda }_{\textup{QCD}}}{\Lambda _{\textup{QCD}}}\simeq \left ( \frac{f}{v} \right )^{2/9}
    \label{eq:Lambda_Lambdatwin}
\end{align}
This scaling would result in a twin nucleon mass not much above $1\GeV$. Thus, a hard $\Z2$ breaking was usually assumed in QCD running couplings to make the twin nucleon heavier. The central point of this work is to show that it is possible to have a $\simeq 1\GeV$ twin nucleon  DM candidate without resorting to hard $\Z2$ breaking. To show that, we can rewrite \eqref{eq:omegaDm_omegaB} to make explicit the dependence on the baryon and twin baryon asymmetries, $Y_{\Delta B}$ and $Y_{DM}$. This results in
\begin{equation}
    \frac{\Omega_{DM}}{\Omega_{B}}=\frac{\widetilde m_p}{m_p}\frac{Y_{\rm DM}}{Y_{\Delta B}}\left(\frac{1-r}{1-\widetilde r}\right),
    \label{ratioden}
\end{equation}
where we defined the baryon and twin baryon fractional asymmetries $r$ and $\widetilde r$ as
\begin{align}
    &r=\frac{n_B}{n_{\overline{B}}}, \hspace{0.5cm} \widetilde r=\frac{\widetilde n_{B}}{\widetilde n_{\overline{B}}}.
\end{align}
Because we assumed that the SM and twin sectors have the same mechanism to generate the asymmetry, the fractional asymmetries are expected to be the same. Then, using either \eqref{eq:Yield_2} or \eqref{eq:Y_nonthermal} and $m_p^{\text{twin}}/m_p\simeq (f/v)^{2/9}$, we can write \eqref{ratioden} as 
\begin{equation}
    \frac{\Omega_{DM}}{\Omega_{B}}=\left( \frac{f}{v}\right)^{2/9} \left(\frac{3 \mathcal{F}_{S}\left ( \Delta_{\widetilde N} \right )+\mathcal{F}_{V}\left ( \Delta_{\widetilde N} \right ) }{ 3 \mathcal{F}_{S}\left ( \Delta_N \right )+\mathcal{F}_{V}\left ( \Delta_N \right ) }\right)\frac{|\widetilde\lambda_{u2}|^2}{|\lambda_{u2}|^2}\left|\frac{\sin\widetilde\phi}{\sin\phi}\right|,
    \label{eq:ratio_DM_to_B}
\end{equation}
where $\Delta_N=M_{N_2}^2/M_{N_1}^2$ and $\Delta_{\widetilde N}=M_{\widetilde N_2}^2/M_{\widetilde N_1}^2$.

\begin{figure}
    \centering
    \includegraphics[scale=0.38]{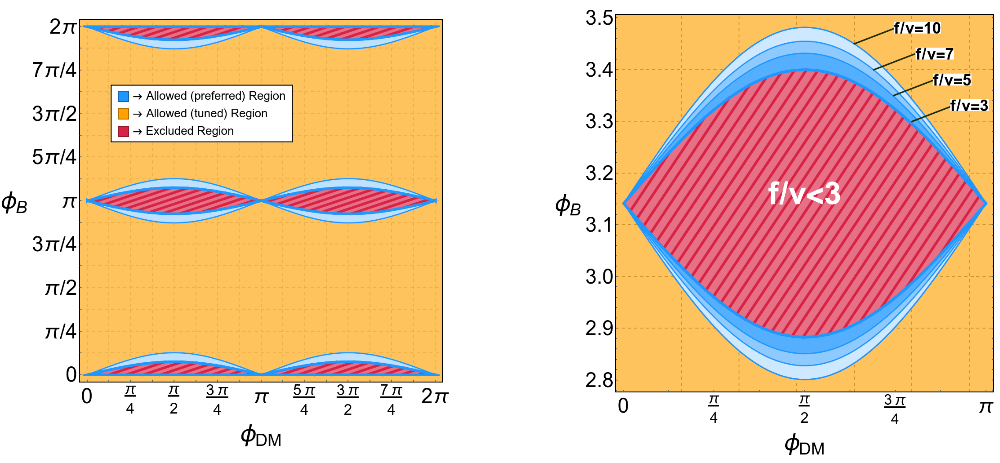}\hspace{1.5cm}
    \caption{Allowed and excluded phases in radians of the neutral fermion $(\phi=\phi_B)$ and twin-fermion $(\widetilde\phi=\phi_{DM})$ coupling to reproduce the observed ratio of the DM abundance to visible matter. Following \eqref{eq:Abundance_ratio_twin}, the excluded region corresponds to a twin sector scale $f>3v$ that is incompatible with measurements of invisible Higgs decays. The preferred region is obtained assuming a tuning of $f<10v$.}
    \label{fig:phases}
\end{figure}

Since we are not interested in introducing hard $\Z2$ breaking, we set $|\lambda_{u2}|=|\widetilde\lambda_{u2}|$. Also, we assume that the mass splittings $\Delta_N$ and $\Delta_{\tilde N}$ are the same in both sectors. In any case, and as already discussed in Section~\ref{sec:Bgen}, the final abundances only depend weakly on the change of these parameters. Finally, because of the $\Z2$ symmetry, the fractional asymmetries in both sectors should be the same. In this way we can rewrite \eqref{eq:ratio_DM_to_B} as
\begin{equation}
   \frac{\Omega_{DM}}{\Omega_{B}} \simeq \left( \frac{f}{v}\right)^{2/9} \left|\frac{\sin\widetilde\phi}{\sin\phi }\right|\simeq 5.
    \label{eq:Abundance_ratio_twin}
\end{equation}
Therefore, it is possible to satisfy the ADM requirement for the DM to baryon abundance if the phases in the visible and twin sectors are misaligned. In Fig.~\ref{fig:phases}, we show the allowed phases needed in order to satisfy relation \eqref{eq:Abundance_ratio_twin}.

As we argued in section \ref{sec:admth}, the misalignment of the phases can be viewed as an IR effect and does not qualify as hard $\Z2$ breaking. One source of misalignment comes from the relative phase between the SM and twin vevs, $\delta$, defined by \eqref{eq:twin_VEV}. Once there is a relative phase difference in the SM and twin masses, the CKM field redefinition introduces different phases on the couplings of \eqref{eq:NdecaysBgen} and \eqref{eq:Twin_lag}.
\begin{alignat}{5}
    \textbf{SM:} & \text{ (Flavor Basis)}&&\quad \lambda_{i\alpha} N_\alpha \overline{X}^a (u_R^{~i})_a \longrightarrow \lambda_{i\alpha} U^{ij}_{u,R} N_\alpha \bar{X}^a (u_R^{~j})_a \quad &&\text{(Mass Basis).}
    \label{eq:Flavor_to_mass_basis1}\\
    \textbf{Twin:} & \text{ (Flavor Basis)}&&\quad \widetilde\lambda_{i\alpha} \widetilde N_\alpha \overline{\widetilde X}^a (\widetilde u_R^{~i})_a \longrightarrow \widetilde\lambda_{i\alpha} e^{i\delta} \widetilde U^{ij}_{u,R} \widetilde N_\alpha \bar{X}^a (u_R^{~j})_a \quad &&\text{(Mass Basis).}
    \label{eq:Flavor_to_mass_basis2}
\end{alignat}

Here, $U_{u,R}$ and $e^{i\delta} \widetilde U_{u,R}$ are the unitary matrix used to diagonalize the Yukawa terms in the SM and twin sector, respectively. Even if the exact $\Z2$ ensures $\lambda_{i\alpha}=\widetilde\lambda_{i\alpha}$ in \eqref{eq:Flavor_to_mass_basis1} and \eqref{eq:Flavor_to_mass_basis2} if the phase is non-zero we expect different imaginary parts of the visible and twin couplings.
\begin{equation}
\Im{\lambda_{i\alpha} U^{ij}_{u,R}}\neq \Im{\widetilde\lambda_{i\alpha} e^{i\delta}\widetilde U^{ij}_{u,R}}.
\end{equation}

We conclude that having different phases is not a hard breaking of the $\Z2$ since it is still a symmetry of the UV theory and does not affect the hierarchy problem in any way. Concerning the UV theory, we are not addressing the specific structure of the twin Higgs model in the UV; we only assume that the $\Z2$ can arise as an exact symmetry at those scales. Then, going to IR scales, the phase misalignment mechanism could have other sources beyond the twin Higgs potential. In general, it is difficult to point out all the sources of phase misalignment since this would imply that we have complete knowledge of the flavor sector of the theory\footnote{Even in the SM, we do not have information on the origin of the CP phase in the CKM matrix for example.} and all the relaxation mechanisms that took place in the thermal evolution of the model. Therefore, we are justified in treating the phases $\phi$ and $\widetilde\phi$ as unknown parameters and scanning for the values that reproduce the DM to baryon ratio as in Fig. \ref{fig:phases}. Once we reproduce the observed abundances of DM and visible matter, as well as the baryon asymmetry, we can study the phenomenological implications of the model and see how it can be constrained or observed in the future. 

\section{Phenomenology}
\label{sec:pheno}

Once we have obtained the observed baryon asymmetry from the twin ADM model, we can study the phenomenological constraints and signals of the model. Usually, the most important constraints on twin Higgs models come from cosmology. In the original implementation of the MTH, a mirror SM copy in a hidden sector induces significant contributions to dark radiation at Cosmic Microwave Background (CMB) and Big Bang Nucleosynthesis (BBN) epochs. If the twin sector had a thermal history similar to the SM, relativistic twin neutrinos and photons would contribute to the total effective number of relativistic degrees of freedom of the universe. 

Many solutions in the literature deal with the potential cosmological problems of Twin Higgs models. One straightforward approach is to decrease the number of relativistic degrees of freedom in the Twin sector with explicit $\Z2$ breaking. As discussed in section \ref{sec:intro}, this strategy is used in the Fraternal Twin Higgs model where only the third generation of fermions are kept in the Twin sector \cite{Craig:2015pha}. This structure is the minimal particle content to address the electroweak Hierarchy Problem. However, it relies on the hard breaking of the $\Z2$ symmetry at the UV scale, which does not apply to this work. Another possibility is an asymmetric reheating that injects more energy into the SM sector than the twin sector. This possibility was considered in \cite{Chacko:2016hvu,Curtin:2021alk,Ireland:2022quc,Koren:2019iuv}. In asymmetric reheating, a massive long-lived particle freezes out from the thermal bath while still relativistic. As the universe expands, they become non-relativistic and decay in both sectors after they decouple. However, in these mechanisms, the massive particle decays are preferentially arranged to decay into the SM sector. As a consequence, the temperature in the visible sector will be bigger than in the Twin sector, alleviating the $\Delta \textup{N}_{\textup{eff}}$ tension. In principle, we could implement asymmetric reheating in the decays of $N_1$ of our model. However, since there are preferential decays to the SM, the numerical predictions we found in the last section would change. Instead, we assume a more straightforward solution to the $\Delta \textup{N}_{\textup{eff}}$ tension and leave other implementations for future work. 

A simple solution to the cosmological problems of the MTH was introduced by \cite{Chacko:2016hvu} and worked out in \cite{Csaki:2017spo,Costa:2020kix,Holst:2023hff}. The idea is to give a large mass to twin neutrinos, making them decouple non-relativistically from the thermal plasma much earlier in cosmic history. Effectively, the twin neutrino contribution to $\Delta \textup{N}_{\textup{eff}}$ is removed. We assume that a seesaw-like mechanism exists and is responsible for generating large twin neutrino masses. The implementation details can be found in previously mentioned literature on the twin neutrinos solution to $\Delta{\rm N}_{\rm eff}$.

Once the twin neutrinos are heavy, our scenario in the MTH has a single viable candidate for DM - twin neutrons. The argument proceeds as in \cite{Farina:2015uea}. Since $\Delta{\rm N}_{\rm eff}$ with heavy twin neutrinos is within experimental bounds, we can keep the twin photon in the spectrum to enforce the $\Z2$ symmetry\footnote{If the SM prediction of $\Delta{\rm N}_{\rm eff}$ remains confirmed with future data, keeping the twin photon in the model can become problematic beyond the $3\sigma$ level. If this becomes the case, one could implement asymmetric reheating to avoid the cosmological bounds or have a massive twin photon as in \cite{Chacko:2019jgi,Harigaya:2019shz}.}. Only twin baryon number $\tilde B$ is generated with no twin Lepton asymmetry. Therefore, all twin leptons can annihilate, and twin electrons are not DM candidates\footnote{If the twin electrons do not annihilate, they could potentially add to the abundance to the point of leading to overclosure.}. The charge neutrality of the universe requires that there is no net production of twin protons since they cannot combine into neutral objects. Therefore, after the twin QCD phase transition a net twin neutron $\tilde n$ number is generated. Finally, twin nucleosynthesis cannot proceed in the presence of heavy twin neutrinos since there are no protons to combine with neutrons. Because the neutron is stable and the only twin relic, we conclude that dark matter is made entirely of $\tilde n$. 

Now that we have established that the MTH DM candidate is the  twin neutron, we can study the direct detection signals. As previously mentioned, we can estimate the mass of the $\tilde n$ to be near that of the visible nucleons, corrected  by the twin sector scale. The precise relation follows from the definition of the QCD scales in both sectors. The leading order contribution to $\Lambda_{\rm QCD}$ arises from the running of the strong couplings coupling,
\begin{equation}
    \alpha_s(Q^2)=\frac{1}{b_0(N_f) \ln\tfrac{Q^2}{\Lambda^2_{\rm QCD}}} 
    \label{eq:run_alphaS}
\end{equation}
 Here, we have defined $b_0(N_f)=33-2 N_f$ and $N_f$ is the number of active quark flavors lighter than the relevant scale $m_f<Q$. Because of the $N_f$ dependence, there will also be an effect on $\Lambda_{\rm QCD}$ due to the quark mass thresholds. In appendix \ref{app:QCD_twinQCD}, we compute the mass-threshold contributions due to integrating out the heavy quark states $f=t,b,c$ to the QCD scale. If we divide the QCD and twin QCD scales, we obtain the following relation
\begin{equation}
    \frac{\widetilde\Lambda_{\rm QCD}}{\Lambda_{\rm QCD}}\sim \left(\frac{\widetilde{y}_t}{y_t}\frac{\widetilde{y}_b}{y_b}\frac{\widetilde{y}_c}{y_c}\right)^{2/27}\left(\frac{f}{v}\right)^{2/9}\exp\left[-\frac{2\pi}{9}\left(\frac{1}{\tilde\alpha_s}-\frac{1}{\alpha_s}\right)\right]
\end{equation}
Assuming there is no hard $\Z2$ breaking, we can set $\tilde{y}_f=y_f$ and $\tilde\alpha_s=\alpha_s$. Finally, there is only a soft $\Z2$ breaking due to the heavier vev of the twin sector, and we recover equation \eqref{eq:Lambda_Lambdatwin},
\begin{align}
    \frac{\tilde{\Lambda }_{\textup{QCD}}}{\Lambda _{\textup{QCD}}}\simeq \left ( \frac{f}{v} \right )^{2/9}.
\end{align}
Since the neutron and twin-neutron masses are proportional to their respective QCD scales, we can write 
\begin{align}
    m_{\rm DM} = m_{\rm \Tilde{n}}=\left(f/v\right)^{2/9} m_{\rm n}.
\end{align}
Then, assuming that $f/v\gtrsim 3$ from the LHC Higgs coupling measurements~\cite{ATLAS:2021vrm,CMS:2022dwd} and $f/v\lesssim 10$ to limit the fine-tuning of the model, we arrive at a rather narrow range of DM mass in this model:
\begin{equation}
    1.2\GeV \lesssim m_{\rm DM} \lesssim 1.6 \GeV.
    \label{eq:mass_range}
\end{equation}

Next, we estimate the nucleon-DM cross-section. The starting point is understanding the halo's local dark matter profile. Because twin dark matter is twin neutrons, its self-interactions should be of the order of the nucleon cross-sections, around $\sim 1cm^2/g$ at energies of a few $\GeV$. Because of this value, we observe that twin DM is within or borderline close to the bounds from small-scale structure formation and merging clusters \cite{Tulin:2017ara,Bullock:2017xww}. While the suppression of small-scale structure could be a signal of this or other similar ADM  models, we leave this part for future work. Several complications are still under debate in the literature regarding the need for suppression in small scales\footnote{The reliability of the collisionless cold dark matter simulations to predict small-scale structure suppression and the role of baryonic feedback are some examples of recent discussions in the literature.}. We therefore assume the twin-neutron self-interaction cross-section satisfies the bound. Assuming this, we then expect twin dark matter to have an approximately uniform distribution within the galaxy halo, allowing for the usual dark matter halo profile and velocity distribution. 

Direct detection of twin dark matter assumes that the two sectors communicate. This communication can occur either through the Higgs portal or other operators at the UV completion scale of the MTH model. Since we are interested in the scattering of nuclei and twin DM at low energies, we can use the effective theory of light quarks and twin quarks. Generically we can write
\begin{align}
    \mathcal{L}_{\rm eff}=\frac{c_{q\Tilde q}^{ij}}{\Lambda^2}(\overline{q_i}\Gamma q_i)(\overline{\Tilde{q}_j}\Tilde\Gamma \Tilde{q}_j),\hspace{1cm} \Gamma,\Tilde\Gamma=\mathds{1}, i\gamma_5,\gamma^\mu, \gamma_5\gamma^\mu,\sigma^{\mu\nu},
    \label{eq:eff_quark}
\end{align}
where, $i=u,d,s$ and $j=\tilde u,\tilde d, \tilde s$. $c_{q\Tilde q}^{ij}$ are the Wilson coefficients of the operator and $\Lambda$ is some scale high compared to  $1 \GeV$. In general, we can write different Lorentz structures, $\Gamma$. However, for our purposes, we are only interested in effective quark operators that generate spin-independent interactions that survive the point-like nucleon approximation. Therefore, we only keep $\Gamma,\Tilde\Gamma=\mathds{1},\gamma^\mu$ since these generate spin independent NR interactions \cite{DelNobile:2021wmp,Fan:2010gt}. 

In the case of \eqref{eq:eff_quark} being generated by the Higgs portal interaction, we have the following scalar 4-fermion operator,
\begin{equation}
    \mathcal{L}_{\rm eff}^{\rm higgs}=\frac{y_i y_j}{m_h^2}\xi (\overline{q_i} q_i)(\overline{\Tilde{q}_j} \Tilde{q}_j),
\end{equation}
where $\xi=v^2/f^2$ and we used the $\Z2$ symmetry to write the twin quark Yukawa coupling $\tilde{y}_j$ to be equal to the visible Yukawa couplings. In this case, we expect this operator to generate a small nucleon cross-section since a double suppression comes from the Yukawa couplings of the light SM and twin quarks.

The other possibility is that the effective operators \eqref{eq:eff_quark} are generated at the MTH cutoff. Considering this case, we can write the two operators that generate spin-independent non-relativistic interactions, a scalar and a vector operators:
\begin{align}
    \mathcal{L}_{\rm eff}^{\Lambda_S}=\frac{c_{S}}{\Lambda_S^2}(\overline{q_i} q_i)(\overline{\widetilde{q}_j} \widetilde{q}_j), \hspace{1cm} \mathcal{L}_{\rm eff}^{\Lambda_V}=\frac{c_{V}}{\Lambda_V^2}(\overline{q_i} \gamma^\mu q_i)(\overline{\widetilde{q}_j} \gamma_\mu \widetilde{q}_j).
\end{align}
Furthermore, we absorb the coefficients of the scalar and vector operators into the definition of the cutoff scales $\Lambda_{S,V}$, effectively setting $c_S=c_V=1$.  

\begin{figure}
    \centering
    \includegraphics[scale=0.3]{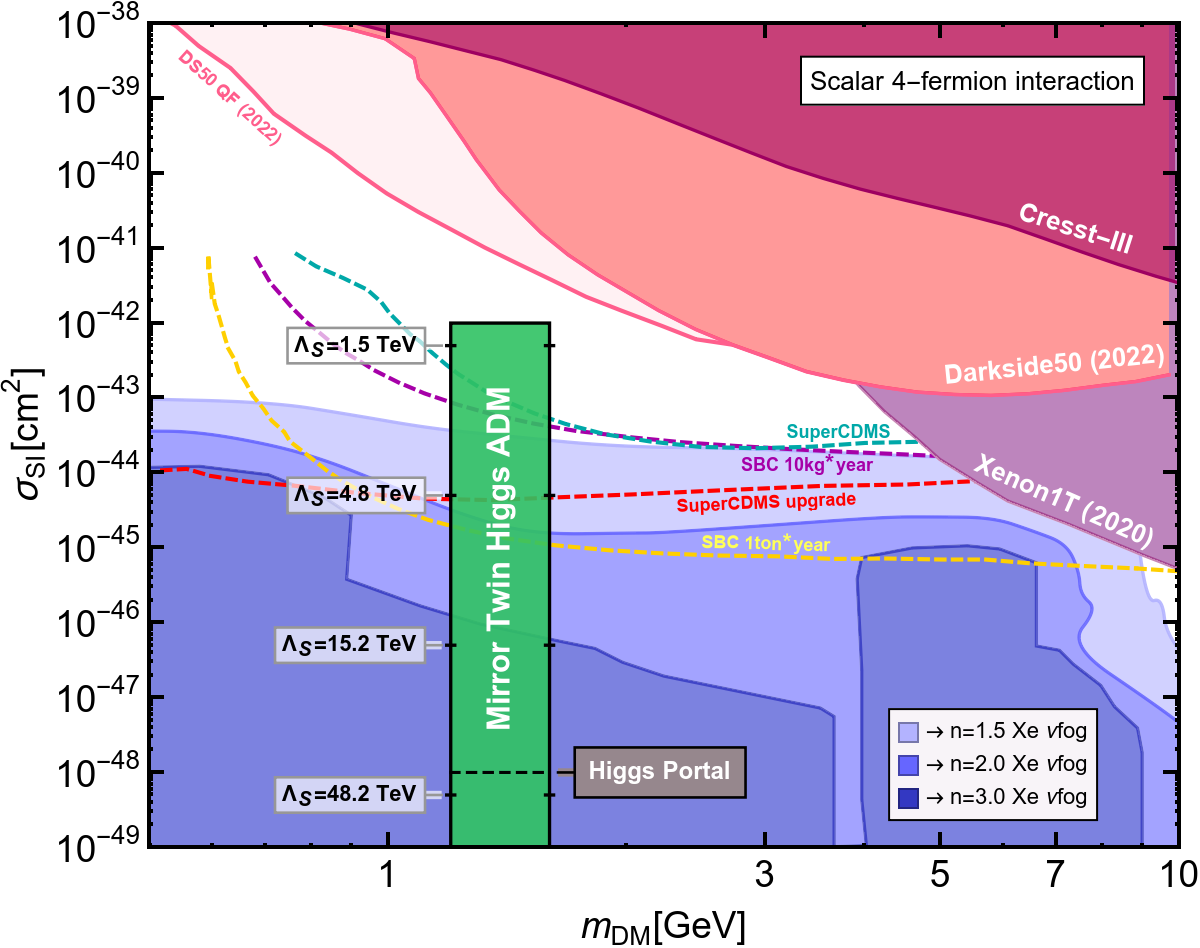}\hspace{0.5cm}
    \caption{Spin-independent cross section parameter space for direct detection of twin dark matter. In this plot, dark matter is made of twin neutrons that interact through the scalar 4-fermion interaction. The green rectangle indicates the twin neutrons' predicted mass and cross-section. Current bounds from DarkSide \cite{DarkSide:2022dhx}, Cresst \cite{PhysRevD.100.102002} and Xenon1T \cite{XENON:2017lvq,XENON:2018voc} are represented by the filled colored regions, including DarkSide quenched analysis (QF) as a solid curve. Dashed lines show future exclusion curves for superCDMS \cite{SuperCDMS:2016wui,SuperCDMS:2022kse} and SBC \cite{Giampa:20211P}, while the blue regions below depict the neutrino fog in the O'hare definition \cite{OHare:2021utq,Akerib:2022ort}.
    \label{fig:DDscalar}}
\end{figure}

The spin-independent cross-section, $\sigma_{SI}$, can be calculated using standard methods as described in appendix \ref{app:SI_cross_section}. For the scalar and vector operators, $\sigma_{SI}$ is given by
\begin{equation}
    \sigma_{SI}^{\rm scalar}=\frac{\mu_{N\tilde n}^2}{\pi}\frac{f_N^2 f_{\Tilde n}^2}{\Lambda_S^4}, \hspace{1.5cm} 
    \sigma_{SI}^{\rm vector}=\frac{\mu_{N\tilde n}^2}{\pi}\frac{b_N^2 b_{\Tilde n}^2}{\Lambda_V^4}.
    \label{eq:SI_cs}
\end{equation}
where $\mu_{N\tilde n}$ is the reduced mass of the twin-neutron and nucleon system, and the zero momentum constants are derived from the form factors as
\begin{align}
    &f_N=\sum\limits_{q} f_{Tq}^{(N)}\simeq 0.3, \hspace{1.2cm} 
    f_{\widetilde n}=\sum\limits_{\widetilde q} f_{T\tilde{q}}^{(\Tilde n)} \simeq 0.3,  \label{eq:fNn}\\
    &b_N=\sum_q F_1^{q,N}(0)=3, \hspace{1cm}
    b_{\widetilde n} =\sum_{\widetilde q} \widetilde{F}_1^{q,N}(0)=3 \label{eq:bNn}.
\end{align}
Notice that the vector form factors are ten times larger than the scalar ones at zero momentum. Since the form factor goes with the fourth power in \eqref{eq:SI_cs}, there will be a significant difference in reach for the scales in the vector and scalar operators.

\begin{figure}
    \centering
    \includegraphics[scale=0.3]{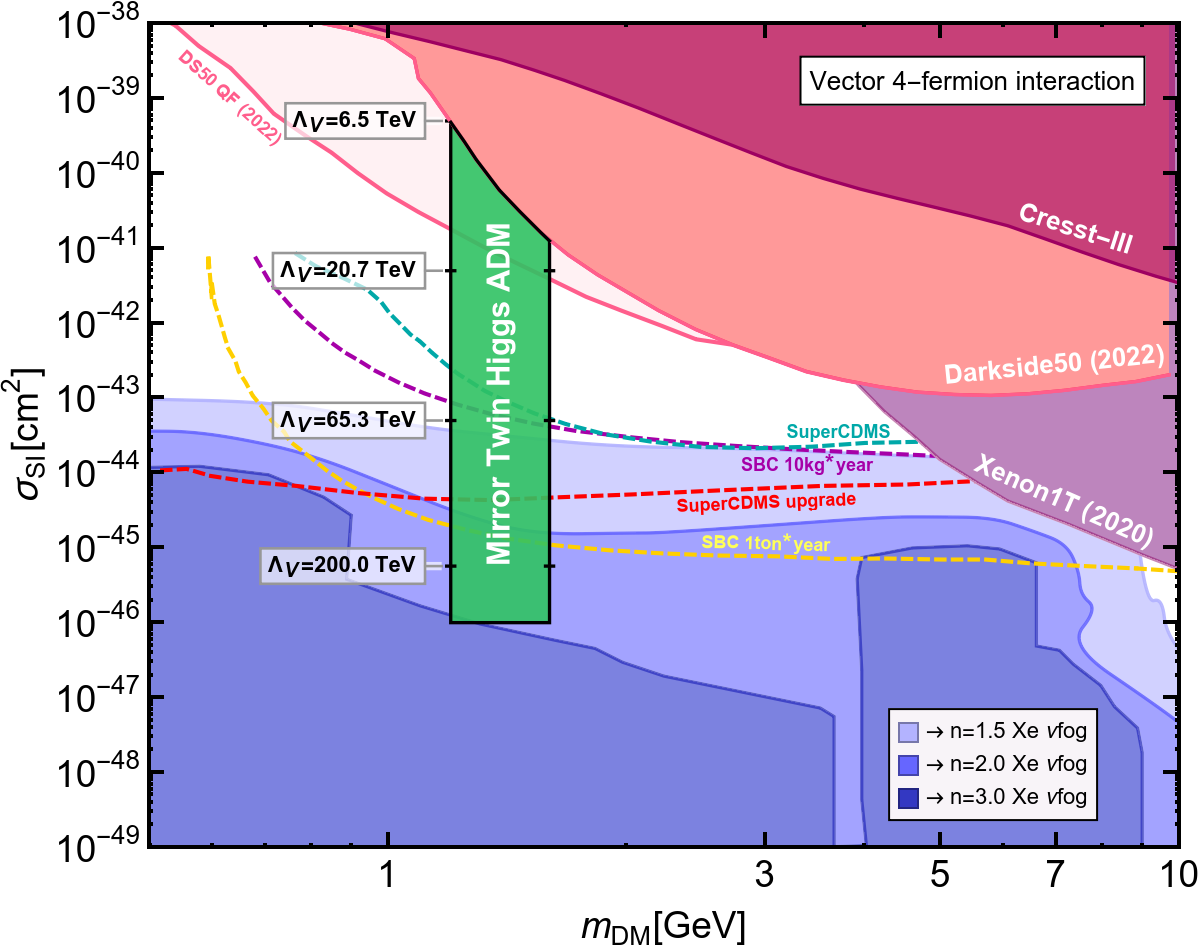}\hspace{0.5cm}
    \caption{Spin-independent cross section parameter space for direct detection of twin dark matter. Dark matter is made of twin neutrons that interact through the vector 4-fermion interaction in this plot. The green rectangle indicates the twin neutrons' predicted mass and cross-section. Current bounds from DarkSide \cite{DarkSide:2022dhx}, Cresst \cite{PhysRevD.100.102002} and Xenon1T \cite{XENON:2017lvq,XENON:2018voc} are represented by the filled colored regions, including DarkSide quenched analysis (QF) as a solid curve. Dashed lines show future exclusion curves for superCDMS \cite{SuperCDMS:2016wui,SuperCDMS:2022kse} and SBC \cite{Giampa:20211P}, while the blue regions below depict the neutrino fog in the O'hare definition \cite{OHare:2021utq,Akerib:2022ort}.
    \label{fig:DDvector}}
\end{figure}

In Figures~\ref{fig:DDscalar} and \ref{fig:DDvector}, we show the spin-independent twin-neutron nucleon scattering cross-section parameter space for the scalar and vector operators, respectively. The green rectangle shows the allowed DM mass given by \eqref{eq:mass_range} and the scale of the operator for each cross-section. The different plots highlight the contrasting reach of the scalar and vector scales, with high $\Lambda_S$ down into the neutrino fog. In the scalar case, the Higgs portal appears at a higher effective scale, around $\Lambda_S \sim 40\TeV$, due to the double suppression of the first generation Yukawa couplings. The filled regions are the current exclusion bounds from Darkside 2022 data \cite{DarkSide:2022dhx}, CRESST-III \cite{PhysRevD.100.102002} and XENON1T \cite{XENON:2017lvq,XENON:2018voc}. In the case of Darkside, nuclear recoils are subject to quenching effects, which cause a reduction in the energy signal due to various mechanisms whose statistics are not fully understood. Because of these effects, \cite{DarkSide:2022dhx} considered two models to bound the quenching effect region where quenching fluctuations are suppressed (NQ) or unsuppressed (QF). NQ corresponds to the filled solid pink region (DarkSide50 2022), and QF is the DS50 QF curve in Figs. \ref{fig:DDscalar} and \ref{fig:DDvector}. While the quenching factor can vary between events, it is typically quantified using calibration sources and simulations. Once these analyses are done, the real exclusion region should lie somewhere in between the NQ and QF curves.

For the neutrino background, we present the Xenon neutrino fog as defined by \cite{OHare:2021utq,Akerib:2022ort}. The index n, the gradient of the DM discovery limit over some exposure measure, labels the different neutrino fog curves and is given by
\begin{equation}
    n=-\left(\frac{d\log \sigma}{d\log N}\right)^{-1} ,
\end{equation}
where $\sigma$ is the discovery limit, and N is the number of events. Given a cross-section experimental sensitivity, this definition means that reducing the sensitivity by a factor of $x$ requires increasing the exposure by $x^n$. Therefore, future experiments can put exclusion bounds inside the neutrino fog region by having sufficient exposure time. The dashed lines correspond to projections by the SuperCDMS \cite{SuperCDMS:2016wui,SuperCDMS:2022kse} and SBC \cite{Giampa:20211P} experiments. A large portion of the parameter space for twin dark matter will likely be probed in the future, especially for the effective vector operators. Due to the smaller scalar form factors, reaching very high cutoff scales is more challenging, and part of the interesting parameter space is down the $n>3$ neutrino fog region. Promising strategies beyond maximizing exposure could be adopted to probe this region. One of these is using the directionality of the neutrino flux to reduce their background. For a review of direct detection prospects below the neutrino fog, we point out to \cite{Akerib:2022ort}.

We can conclude from the figures above that the interesting scales for the UV completion of the MTH, typically of the order of $10~$TeV, are beginning to be probed by the Darkside collaboration. This is clearly the case for the vector operator (Figure~\ref{fig:DDvector}). On the other hand, for the scalar operator (Figure~\ref{fig:DDscalar}) the suppressed sensitivity resulting from the smaller zero momentum constants in (\ref{eq:fNn}) puts this interesting UV completion scale under the neutrino fog, making its detection more challenging. Additionally, the Higgs portal should be always present in the MTH independently of the UV completion. Therefore, reaching the Twin Higgs portal cross section has the potential of excluding or confirming the model. However, due to the double first-generation Yukawa coupling suppression, the signal for direct detection goes deep into the neutrino fog, with difficult experimental prospects. In any case, we see that the direct exploration of the parameter space of this ADM scenario of the MTH is becoming feasible in current and future experiments. 

To finish this section, we briefly comment on other possibilities for the phenomenology of the presented model. First, bounds from neutrons oscillation experiments do not apply here since the interactions of $X$ and quarks in \eqref{eq:NdecaysBgen} is anti-symmetric in flavor. Additionally, the charged $X$ production could be explored at the LHC. This paper assumes that $X$ is heavy enough to be out of reach by collider experiment. However, we are pursuing the collider phenomenology of this low-temperature baryogenesis scenario in a forthcoming publication.

\section{Conclusions}
\label{sec:conc}

The primary focus of this paper is to present a Mirror Twin Higgs implementation of asymmetric dark matter in order to show that  there is no need to introduce a hard $\Z2$ breaking  in order to have a consistent dark matter candidate. We focused on adding a particular baryogenesis model to the visible sector of the MTH that successfully generates the baryon asymmetry at low temperatures after sphaleron decoupling. The model relies on the out-of-equilibrium decays of a neutral fermion that violates baryon number and CP symmetry. We showed that the observed baryon asymmetry can be correctly obtained, assuming that the neutral fermion is produced in the early universe, independently of weather this production is thermal or non-thermal. 

We emphasize that any specific model of baryogenesis that is then mirrored to the MTH through the $\Z2$ symmetry will result in the same conclusion: no hard $\Z2$ breaking is necessary as long as the needed $\Z2$ breaking is in the relative phases of the couplings appearing in the CP violation asymmetries in the visible and twin sectors, which are responsible for the baryon and DM abundances. This generality was first mentioned in section~\ref{sec:admth}, in the paragraph prior to eq.~(\ref{eq:Abundance_ratio_5}), where it is clear that as long as the CP violating phases as misaligned by order one factors, it is possible to obtain the correct DM and baryon number abundances without making use of {\em hard} $\Z2$ breaking in the running of the twin QCD coupling, as it was thought to be required in Ref.~\cite{Farina:2015uea}. In the specific model we use this translates in (\ref{eq:Abundance_ratio_twin}), which has the same scaling with the phases introduced in sections~\ref{sec:admth} and \ref{sec:baryonadm}.

The $\Z2$ symmetry of the MTH model extends the cosmological mechanism responsible for baryogenesis to the twin sector. This mirroring gives rise to an abundance of asymmetric dark matter predominantly composed of twin neutrons. Misalignment of the complex phases between the visible and twin sectors make possible a dark matter abundance consistent with the observed value of $\Omega_{\rm DM} \simeq 5 \Omega_B$. This phase misalignment could arise solely as an IR effect, ensuring that no hard breaking $\Z2$ needs to be introduced. A simple example is vacuum misalignment between the two scalar sectors, leading to different phases in their couplings entering the CP asymmetries. This, as well as similar misalignments in the phases of fields entering the singlet couplings in both sectors, can be IR effects and therefore thought of as soft $\Z2$ breaking,  Preserving the UV $\Z2$ symmetry of the twin Higgs model is a desirable feature concerning the electroweak stability of the theory, and its protection guarantees that the solution to the little hierarchy problem remains unspoiled, without the need of further tunings.

As mentioned above, it is possible to reach the same final DM abundance with different baryogenesis implementations. Because of this, our results are not limited to the specific baryon asymmetry mechanism used in section \ref{sec:baryonadm}. Once a visible baryon asymmetry is achieved, the $\Z2$ symmetry and phase misalignment are enough to reproduce the abundance ratio between DM and baryons. Consequently, the mechanism exhibits the potential for generalization to alternative baryogenesis models, such as high-scale leptogenesis or electroweak baryogenesis. These extensions can be explored in future developments of ADM in the MTH framework. 

Regardless of the chosen baryogenesis model, implementing ADM in the MTH model without hard $\Z2$ breaking predicts that dark matter consists mainly of twin neutrons with masses ranging from $1.2 ~\GeV$ to $1.6 ~\GeV$. A significant part of the parameter space is probed assuming an effective interaction of the light quarks and twin quarks. Part of the parameter space is excluded by the data from the Crest-III and Darkside-50 experiments. Promisingly, future experiments such as SuperCDMS and SBC can probe higher effective scales beyond the TeV range. Furthermore, direct detection experiments below the neutrino fog hold significant potential for uncovering the nature of twin asymmetric dark matter. We conclude that the mirror twin Higgs model is a well-motivated BSM approach to address the electroweak stability, the nature of DM, as well as the  origin of the baryon asymmetry with the same core concepts. The model presents a compelling candidate for dark matter in the $\GeV$ range, requiring extensive exploration through future DM detection experiments.

\vspace{1cm}
\hrule 
\vspace{0.4cm}

\acknowledgments
The authors thank Ivone Albuquerque, Nicolás Bernal, Chee Sheng Fong and Seth Koren for helpful discussions. 
They also acknowledge the support of FAPESP grants 2019/04837-9 and 2021/02757-8, and CAPES 88887.816450/2023-00. 

\appendix
\section{QCD and twin QCD scales}
\label{app:QCD_twinQCD}
In this appendix we derive the leading order relationship between the SM QCD and twin QCD scales we used throughout the text,
\begin{equation}
    \frac{\widetilde\Lambda_{QCD}}{\Lambda_{QCD}}= \left(\frac{f}{v}\right)^{2/9}.
    \label{eq:twinQCD_QCD}
\end{equation}

Following \cite{Ellis:1996mzs}, the derivation makes use of the quark-mass threshold contributions to $\Lambda_{QCD}$. At leading order, the running coupling $\alpha(Q^2,N_f)$ can be written for momentum greater than the top quark as
\begin{equation}
    \alpha_s(Q^2,6)=\frac{1}{b(6)\log Q^2/\Lambda_{UV}^2}, \hspace{1cm} Q^2>m_t^2.
\end{equation}
Here, $N_f=6$ is the number of active quark flavors at high energies and $b(N_f)=33-2N_f$. Crucially, the UV QCD scale defined in this relation is the same between the visible and the twin sector.

The first quark threshold correction appears when we integrate out the top-quark. The coupling below the top quark mass can be written as
\begin{equation}
    \alpha_s(Q^2,5)=\frac{1}{b(5)\log Q^2/\Lambda_{UV}^2}+c, \hspace{1cm} Q^2>m_b^2,
\end{equation}
where $c$ is a constant fixed by requiring the matching between the theory with six and five quarks at the top-quark mass scale, $\alpha_s(m_t,6)=\alpha_s(m_t,5)$. Calculating $c$ we arrive at
\begin{equation}
    \frac{1}{\alpha_s(Q^2,5)}=b(5)\log\frac{Q^2}{m_t^2}+b(6)\log\frac{m_b^2}{\Lambda_5^2},
    \label{eq:alpha6to5}
\end{equation}
 However, $\alpha_s(Q^2,5)$ also defines the QCD scale for the theory with only 5 active quarks, $\Lambda_5$.
\begin{equation}
    \frac{1}{\alpha_s(Q^2,5)}=b(5)\log \frac{Q^2}{\Lambda_5^2}, \hspace{1cm} Q^2>m_b^2.
    \label{eq:alpha5}
\end{equation}
Comparing \eqref{eq:alpha6to5} to \eqref{eq:alpha5}, we arrive at a relation for the 5 quark-flavors QCD scale.
\begin{equation}
    \Lambda_5=\Lambda_{UV} \frac{m_t^{1-b(6)/b(5)}}{\Lambda_{UV}^{1-b(6)/b(5)}}.
\end{equation}
 We can do the same procedure to obtain the quark-mass threshold contributions to the QCD up to the charm-quark. Beyond this point, the theory becomes strongly interacting and we cannot perturbatively integrate out the light quark-flavors since they are below the QCD scale. Thus, the definition for the QCD scale includes threshold contributions from the three heavy states, the top, bottom and charm quarks. We can them write
\begin{equation}
    \Lambda_{QCD}\equiv \Lambda_3 = \Lambda_{UV}^{b(6)/b(3)} m_t^{\left(1-\tfrac{b(6)}{b(5)}\right)\tfrac{b(5)}{b(3)}} m_b^{\left(1-\tfrac{b(5)}{b(4)}\right)\tfrac{b(4)}{b(3)}} m_c^{\left(1-\tfrac{b(4)}{b(3)}\right)}.
\end{equation}
Substituting the values of $b(N_f)$ and using that the mass-Yukawa relation $m_q=y_q v$, we obtain
\begin{equation}
    \Lambda_{QCD}=\Lambda_{UV}^{7/9} y_t^{2/27} y_b^{2/27} y_c^{2/27} v^{2/9}.
    \label{eq:QCD}
\end{equation}
Similarly, for the twin QCD scale we can write
\begin{equation}
    \widetilde \Lambda_{QCD}=\Lambda_{UV}^{7/9} \widetilde y_t^{2/27} \widetilde y_b^{2/27} \widetilde y_c^{2/27} f^{2/9}.
    \label{eq:twinQCD}
\end{equation}
Since there is no $\Z2$ breaking in the model, we can write the twin Yukawa couplings as $\widetilde y_q =y_q$. Finally, dividing \eqref{eq:twinQCD} by \eqref{eq:QCD} we obtain the proposed relation,
\begin{equation}
    \frac{\widetilde\Lambda_{QCD}}{\Lambda_{QCD}}= \left(\frac{f}{v}\right)^{2/9}.
\end{equation}

\section{Spin-Independent Cross-section of Twin DM}
\label{app:SI_cross_section}
Now, to find the cross-section, we compute the nucleon-DM scattering matrix. To first order in the perturbative expansion, we have the following nucleon amplitude
\begin{align}
    \mathcal{M}_{\rm N} = \langle \widetilde{n'} N'|\mathcal{L}_{\rm eff}|\widetilde{n} N\rangle = \frac{1}{\Lambda^2} \langle N'| \overline{q_i}\Gamma q_i |N \rangle \langle \widetilde{n}'| \overline{\widetilde{q}_j} \Gamma \widetilde{q}_j |\tilde n \rangle.
\end{align}
We use $N'$ and $\Tilde{n}'$ to denote the nucleon and twin-neutron final states, respectively. The nucleon-spinor bilinears can parameterize each matrix element,
\begin{equation}
    \langle N'| \overline{q_i}\Gamma q_i |N \rangle = \sum_{i=u,d,s} F_i^{(N)}(q^2) \,\overline{u}_{N'} \Gamma u_N,
\end{equation}
where $F_l^{(N)}(q^2)$ are the hadronic form factors associated to the nucleons $N=p,n$. For direct detection of twin DM, it is sufficient to use the hadronic form factors at zero transferred momentum since their variation is negligible compared to the recoil energies considered. In this limit, we can relate the scalar form factors with the fraction of the nucleon mass carried by the light quarks,
\begin{equation}
    \langle N| \overline{q}q |N\rangle =\frac{m_N}{m_q}F_i^{(N)}(0)\overline{u}_N u_N = f_{Tq}^{(N)}\overline{u}_N u_N.
\end{equation}
The vector form factors are related to the conserved flavor singlet vector current associated with the baryon number. 
\begin{equation}
    \langle N| \overline{q}\gamma^\mu q |N \rangle = F_i^{q,N}(0)\overline{u}_N \gamma^\mu u_N
\end{equation}
The form factors at zero momentum can be obtained by perturbative and lattice calculations or by experiment. We used the values from \cite{Gupta:2018lvp,Hoferichter:2015dsa,Ellis:2018dmb,Shifman:1978zn}.

Finally, we can calculate the spin-independent cross-section.
\begin{equation}
    \sigma_{SI}^{\rm scalar}=\frac{\mu_{N\tilde n}^2}{\pi}\frac{f_N^2 f_{\Tilde n}^2}{\Lambda_S^4}, \hspace{1.5cm} 
    \sigma_{SI}^{\rm vector}=\frac{\mu_{N\tilde n}^2}{\pi}\frac{b_N^2 b_{\Tilde n}^2}{\Lambda_V^4}.
    \label{eq:SI}
\end{equation}
where $\mu_{N\tilde n}$ is the reduced mass of twin-neutron and nucleon, and we have defined the constants
\begin{align}
    &f_N=\sum\limits_{q} f_{Tq}^{(N)}\simeq 0.3, \hspace{1.2cm} 
    f_{\widetilde n}=\sum\limits_{\widetilde q} f_{T\tilde{q}}^{(\Tilde n)} \simeq 0.3, \\
    &b_N=\sum_q F_1^{q,N}(0)=3, \hspace{1cm}
    b_{\widetilde n} =\sum_{\widetilde q} \widetilde{F}_1^{q,N}(0)=3.
\end{align}
Notice that the vector form factors are ten times larger than the scalar ones at zero momentum. Since the form factor goes with the fourth power in \eqref{eq:SI}, there will be a significant difference in reach for the vector and scalar probes to the scale of the operator.

\bibliography{ATDM}
\bibliographystyle{JHEP}

\end{document}